\documentclass[]{article}

\usepackage[top = 0.75in, right = 0.75in, left = 0.75in, bottom = 0.75in]{geometry}
\usepackage{authblk}
\usepackage{siunitx}
\usepackage{amsmath}
\usepackage{mathtools}
\usepackage{mathrsfs}
\usepackage{amssymb}
\usepackage{graphicx}
\usepackage{textcomp}
\usepackage{epstopdf}
\usepackage{afterpage}
\usepackage{tabulary}
\usepackage{rotating}
\usepackage{multirow}
\usepackage{longtable}
\usepackage{amsthm}
\usepackage{stmaryrd}
\usepackage{url}
\usepackage{float}
\usepackage{tablefootnote}

\title{Social Hierarchy-based Distributed Economic Model Predictive Control of Floating Offshore Wind Farms}
\author[1]{Ali C. Kheirabadi}
\author[1]{Ryozo Nagamune}
\affil{The University of British Columbia, Vancouver Campus, 2054-6250 Applied Science Lane, Vancouver, BC Canada V6T 1Z4}

\begin{document}

\maketitle

\begin{abstract}
	
This paper implements a recently developed social hierarchy-based distributed economic model predictive control (DEMPC) algorithm in floating offshore wind farms for the purpose of power maximization. The controller achieves this objective using the concept of \textit{yaw and induction-based turbine repositioning}~(YITuR), which minimizes the overlap areas between adjacent floating wind turbine rotors in real-time to minimize the wake effect. Floating wind farm dynamics and performance are predicted numerically using FOWFSim-Dyn. To ensure fast decision-making by the DEMPC algorithm, feed-forward neural networks are used to estimate floating wind turbine dynamics during the process of dynamic optimization. For simulated wind farms with layouts ranging from $1 \times 2$ to $1 \times 5$, an increase of 20\,\% in energy production is predicted when using YITuR instead of greedy operation. Increased variability in wind speed and direction is also studied and is shown to diminish controller performance due to rising errors in neural network predictions.

\end{abstract}

\section{Introduction}

This paper outlines the implementation of distributed economic model predictive control (DEMPC) for the purpose of power maximization in floating offshore wind farms. This introductory section provides an overview of the control application along with a description of the DEMPC algorithm.

\subsection{Application}

Wind turbines that are aligned with the free stream wind suffer from a phenomenon termed the \textit{wake effect}. Viscous interaction along the blades of an upstream machine generates a region downstream that is referred to as a \textit{wake} in which the mean wind speed is reduced in comparison to that of the surrounding wind field. Any turbine that is located downstream and aligned with this wake encounters a lower average wind velocity relative to its upstream neighbor, and may thus operate with up to a 60\,\% reduction in power production~\cite{Nilsson2015}.

One approach to mitigating the wake effect is \textit{wind farm control}, which entails the operation of individual turbine actuators in a manner such that power production from the collective is increased. Two conventional wind farm control techniques have been widely investigated in the context of fixed-foundation wind farms; power de-rating, which involves sacrificing power production from upstream turbines to increase the mean wind speeds to which downstream machines are exposed; and wake steering, which requires operating upstream rotors with deliberate yaw misalignment to deflect their generated wakes in the crosswind direction. The reader may refer to our comprehensive review article~\cite{Kheirabadi2019a} for further information on the wake effect and wind farm control.

\begin{figure}
	\centering
	\includegraphics[width=4.5in]{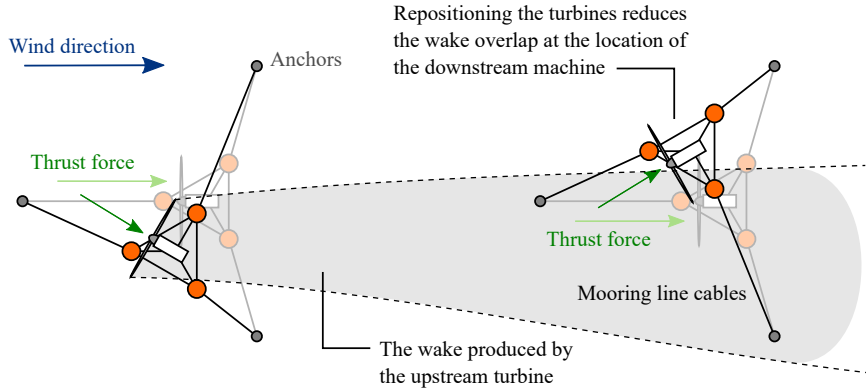}
	\caption{Schematic describing the use of aerodynamic thrust forces to passively relocate floating wind turbines in real-time.} \label{Figure - YITuR operation}
\end{figure}

Floating wind farms render possible a third control strategy for power maximization. Namely, the mooring systems that anchor floating platforms to the seabed permit limited mobility of the platforms along the ocean surface. By operating a wind turbine with deliberate yaw misalignment, the direction of the net aerodynamic thrust force acting on its rotor may be altered, hence shifting its position along the ocean surface. As shown in Fig.~\ref{Figure - YITuR operation}, this mobility may be exploited to reduce the overlap area between the wake generated by an upstream machine and the rotor of a downstream turbine. This wind farm control strategy is termed \textit{yaw and induction-based turbine repositioning} (YITuR). 

Our previous work~\cite{Kheirabadi2020} presented a steady-state model for predicting platform displacements and power production in floating offshore wind farms. Additionally, this work performed a stationary optimization study to evaluate the potential benefits of YITuR under steady-state conditions. To this end, we reported efficiency gains ranging from 21.8~to 42.7\,\% for wind farm sizes spanning from $2 \times 2$ to $7 \times 7$ provided that the mooring lines were long enough to permit crosswind relocation of roughly one rotor radius. Given this potential for increasing energy production, the aim of the current work is to transcend a steady-state analysis and to implement YITuR in real-time using optimal control techniques.

\subsection{Controller}

At any given sampling time, model predictive control (MPC) utilizes a mathematical model of a plant to solve a dynamic optimization problem over a finite prediction horizon. The output of this process is a sequence of optimal control inputs that minimize or maximize a prescribed cost function over the prediction horizon while satisfying constraints. Solely the first step in this sequence is then implemented, at which point, based on updated state and disturbance measurements, the dynamic optimization problem is recomputed at the next sampling time. More detail on MPC is available in review articles by Mayne~\cite{Mayne2000} and Mayne \textit{et al.}~\cite{Mayne2014}.

The turbine repositioning wind farm control problem addressed in this paper does not simply involve set-point stabilization; the control objective is to minimize overlap areas between the rotors of adjacent floating wind turbines. Given this property, economic model predictive control (EMPC) is a prime candidate for automation. EMPC computes optimal control actions by minimizing a \textit{generalized cost function} that is a quantifier of plant performance over a finite prediction horizon. In comparison, traditional MPC minimizes a positive-definite cost function that simply penalizes deviation from a prescribed set-point. More detail on EMPC is available in the review article by Ellis \textit{et al.}~\cite{Ellis2014}.

In order to promote scalability (\textit{i.e.} computation time must not be affected by the size of the plant), it is necessary to distribute the optimal control problem among the various operational turbines within the plant; hence the employment of DEMPC in our approach. DEMPC eliminates the computational burden encountered by centralized EMPC as each agent (\textit{i.e.} floating wind turbine) computes its own optimal control input sequence. In order to reach an optimal decision, every agent must make assumptions as to the state and control input trajectories of neighbors with which it is coupled. A major component of any DEMPC scheme is therefore the coordination algorithm that is used to update the assumptions that each agent holds regarding the future operation of its neighbors. More detail on DEMPC is available in the review article by M\"uller and Allg\"ower~\cite{Muller2017}.

The turbine repositioning problem possesses two properties that pose challenges in the implementation of DEMPC. The first is that objective functions are non-convex. There exist multiple platform displacement paths that reduce the overlap areas between the rotors of adjacent floating turbines. In standard parallel DEMPC algorithms, convergence of agent decisions cannot be guaranteed in the presence of non-convex cost functions~\cite{Liu2012a}; more elaborate sequential~\cite{Kuwata2007,Richards2007}, negotiation-based~\cite{Maestre2011,Muller2012a,Stewart2011}, and group-based~\cite{Liu2019a,Pannek2013,Asadi2018} coordination algorithms are therefore required to establish convergence. These algorithms suffer from scalability issues however~\cite{Kheirabadi2020a}. The second property that poses a challenge is that steady-state terminal set-points are not known \textit{a priori} in the current application. In brief, terminal set-points are necessary for achieving stability in EMPC as they ensure that the computed optimal control actions always lead the system to a stable steady-state~\cite{Ellis2014}. Existing DEMPC methods either assume that these set-points are known \textit{a priori}~\cite{Wang2017a}, are determined using centralized optimization~\cite{Lee2011,Lee2012,Driessen2012a,Chen2012,Albalawi2017a,Wolf2012}, or are computed in a distributed manner assuming convex cost functions~\cite{Kohler2018b}. Our previous work~\cite{Kheirabadi2020a} overcame these drawbacks and presented a DEMPC algorithm that used the concept of social hierarchies to guarantee convergence and stability in the presence of non-convex cost functions and unknown terminal set-points in a truly distributed and scalable manner (\textit{i.e.} no centralized functionalities were required).

\subsection{Contributions}

The main contribution of this paper is the first implementation of a social hierarchy-based DEMPC algorithm for the purpose of power maximization via YITuR in floating offshore wind farms. A secondary contribution includes the first use of feed-forward artificial neural networks to generate a computationally inexpensive control-oriented model of floating wind turbine dynamics. These neural networks serve as the mathematical model used in the dynamic optimization process of the DEMPC algorithm. The final contribution is the first demonstration of the benefits of YITuR implemented in real-time under time-varying wind conditions.

\subsection{Paper organization}

The paper is organized as follows: Section~\ref{Section - Wind farm simulator} describes FOWFSim-Dyn, which is a simulator used to predict platform motion, wake aerodynamics, and wind farm power production; Section~\ref{Section - Controller design} discusses the social hierarchy-based DEMPC objective and algorithm, the use of feed-forward neural networks to approximate turbine dynamics, and the impacts of disturbances and model uncertainty on controller performance; Section~\ref{Section - Results and discussion} provides simulation results of controller implementation corresponding to different wind conditions and turbine array sizes; and Section~\ref{Section - Conclusions} concludes the paper with summaries of major findings and recommendations for future research.

\section{Simulation tool - FOWFSim-Dyn} \label{Section - Wind farm simulator}

\begin{figure}
	\centering
	\includegraphics[width=3in]{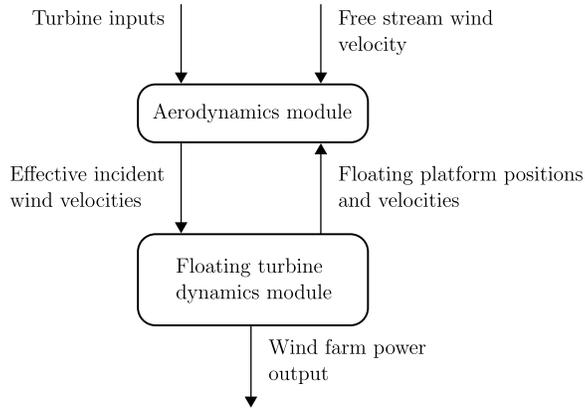}
	\caption{Block diagram showing the computation modules of FOWFSim-Dyn along with information transfer routes.} \label{Figure - FOWFSim block diagram}
\end{figure}

This section briefly describes FOWFSim-Dym, which is a tool used to simulate floating offshore wind farms in the current work. Full details of this software are available in our previous work~\cite{Kheirabadi2020b}. A system diagram of FOWFSim-Dyn is shown in Fig.~\ref{Figure - FOWFSim block diagram}. Two high-level modules are tasked with predicting floating platform motion coupled with wake aerodynamics. The following subsections delineate each of these modules.

\subsection{Floating turbine dynamics module}

\begin{figure}
	\centering
	\includegraphics[width=3.5in]{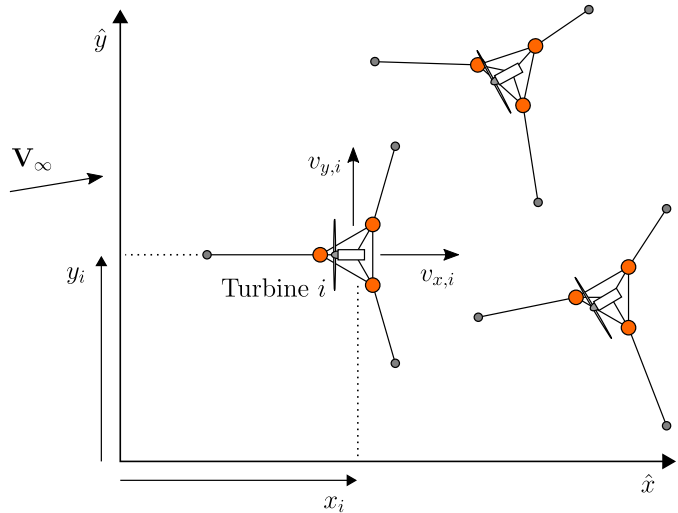}
	\caption{Schematic of the general floating offshore wind farm used in FOWFSim-Dyn.} \label{Figure - Wind farm description}
\end{figure}

The purpose of the floating turbine dynamics module is to predict the positions, velocities, and power outputs of all floating wind turbines within the wind farm. As shown in Fig.~\ref{Figure - Wind farm description}, a floating wind farm is treated as system of particles distributed along the two-dimensional ocean surface. The predominant free stream wind direction is aligned with the global $\hat{x}$ axis; however, FOWFSim-Dyn is capable of simulating fluctuations in the free stream wind direction relative to the $\hat{x}$ axis. The wind velocity vector is denoted as $\mathbf{V}_\infty$.

The variables $x_i$ and $y_i$ describe the positions of turbine~$i$ along the $\hat{x}$ and $\hat{y}$ axes over time. The velocities of the platform of turbine~$i$ in these directions are denoted by $v_{x,i}$ and $v_{y,i}$. Finally, turbine~$i$ possesses two inputs; the rotor's axial induction factor $a_i$ and its yaw angle $\gamma_i$ measured positive counter-clockwise relative to the $\hat{x}$ axis.

The motions of the floating platforms are determined by summing the aerodynamic thrust forces, hydrodynamic drag and added mass forces, and mooring line tensions acting on each turbine. Hydrodynamic drag and added mass forces are calculated using Morison's equation~\cite{Robertson2014}, wherein the drag and added mass contributions of all submerged components are summed to provide the total hydrodynamic force. Mooring line tensions are obtained by solving the problem of a static catenary that is either partially resting along or fully lifted above the seabed~\cite{Kheirabadi2020}.

Aerodynamic thrust forces are estimated by treating wind turbine rotors as actuator discs and computing thrust coefficients according to vortex cylinder theory~\cite{Burton2011}. Computation of aerodynamic thrust forces also requires estimates of the effective wind velocity that is incident upon the rotors of  all turbines. The wake aerodynamics module addresses this requirement.

\subsection{Aerodynamics module}

The wake aerodynamics module predicts characteristics of the wakes generated by each turbine over time. These characteristics include the wake centerline position, wake diameter, and the average wake velocities in the $\hat{x}$ and $\hat{y}$ directions. The evolution of these properties over space and time is captured using one-dimensional momentum conservation with an assumed wake expansion rate.

The boundary conditions to this problem include the initial wake velocities at the location of the turbine that generates each wake. The initial wake velocities in the downwind and crosswind direction are estimated using momentum balance methods proposed by Bastankhah \textit{et al.}~\cite{Bastankhah2016} and Jim\'enez \textit{et al.}~\cite{Jimenez2009}.

The effective wind velocity that is incident upon the rotor of a downstream turbine is computed using the principle of equivalent kinetic power deficit~\cite{Katic1986}. Namely, the net kinetic power deficit at any downstream location is assumed to equate to the sum of kinetic power deficits contributed by all upstream wakes. Further, wake velocity profiles are modeled using Gaussian distributions as per experimental results reported by Bastankhah \textit{et al.}~\cite{Bastankhah2016}.

\section{Controller design} \label{Section - Controller design}

This section first describes the optimal control objective for maximizing power production in floating offshore wind farms. Then, the social hierarchy-based DEMPC algorithm~\cite{Kheirabadi2020a} is briefly described. Next, the use of feed-forward neural networks for estimating floating platform motion during dynamic optimization processes is discussed. These neural networks treat aerodynamic coupling between turbines as disturbances. The current section therefore also discusses the impact of disturbance on the stability and performance of EMPC.

\subsection{Controller objective functions}

Simply put, the objective of DEMPC is to indirectly maximize power production by minimizing the overlap areas between the rotors of adjacent turbines. Power production is not directly maximized since this approach would require the identification of functions that estimate the power output of each turbine using some machine learning scheme. Identification of such functions is challenging since, due to the transport phenomenon, the power produced by any turbine is strongly influenced by the \textit{history} of the states and inputs of its upstream neighbors. In other words, the states and inputs that upstream neighbors would have possessed in the past influence the power produced by a downstream turbine at the present moment. Identification of a power function is therefore deferred to future work who's primary focus is dynamic system identification of floating offshore wind farms. In the current work, the indirect approach of rotor area overlap minimization is taken.

The resulting dynamic optimization problem to be solved by the local EMPC function of turbine~$i$ takes the following form:
\begin{equation} \label{Equation - Dynamic optimization problem}
\min_{\overline{\mathbf{x}}_i,\overline{\mathbf{u}}_i} \sum_{k = 0}^{H - 1} \left\{ \Delta\mathbf{u}_{i,k}^\mathrm{T} \mathbf{Q} \Delta\mathbf{u}_{i,k} + \sum_{j \in \mathcal{N}_i} \left[\frac{1}{\left| \mathcal{N}_i \right|} A_{\mathrm{OL}, i \rightarrow j}(\mathbf{x}_{i,k},\mathbf{x}_{j,k}) + \Delta\mathbf{u}_{j,k}^\mathrm{T} \mathbf{Q} \Delta\mathbf{u}_{j,k} \right] \right\},
\end{equation}
subject to
\begin{subequations}\label{Equation - Dynamic constraints}
	\begin{eqnarray}
	\mathbf{x}_{i,0} & = & \mathbf{x}_i, \label{Constraint - Initial conditions}\\
	\mathbf{x}_{i,k + 1} & = & \mathbf{f}_i(\mathbf{x}_{i,k},\mathbf{u}_{i,k}), \label{Constraint - Dynamic model}\\
	a_{i,k} & = & \frac{1}{3}, \label{Constraint - Axial induction factor}\\
	\left| \gamma_{i,k} \right| & \leq & 10\,\si{deg}, \label{Constraint - Yaw angle}\\
	\mathbf{x}_{i,H} & = & \mathbf{x}_{i,s}, \label{Constraint - Terminal set-point}
	\end{eqnarray}
\end{subequations}
where $H$ is the length of the prediction horizon, and $k$ is the time-step number along the prediction horizon. The vector $\mathbf{x}_i$ contains the most recent state measurements of turbine~$i$ as follows:
\begin{equation}
\mathbf{x}_i \coloneqq
\begin{bmatrix}
x_i & y_i & v_{x,i} & v_{y,i}
\end{bmatrix}^\mathrm{T},
\end{equation}
and the vectors $\mathbf{x}_{i,k}$ and $\mathbf{u}_{i,k}$ contain the candidate states and inputs of turbine~$i$ at time-step number $k$ along the prediction horizon as follows:
\begin{eqnarray}
\mathbf{x}_{i,k} & \coloneqq &
\begin{bmatrix}
x_{i,k} & y_{i,k} & v_{x,i,k} & v_{y,i,k}
\end{bmatrix}^\mathrm{T},\\
\mathbf{u}_{i,k} & \coloneqq &
\begin{bmatrix}
a_{i,k} & \gamma_{i,k}
\end{bmatrix}^\mathrm{T}.
\end{eqnarray}
The terms $\overline{\mathbf{x}}_i \coloneqq \left( \mathbf{x}_{i,0}, \mathbf{x}_{i,1}, \cdots, \mathbf{x}_{i,H} \right)$ and $\overline{\mathbf{u}}_i \coloneqq \left( \mathbf{u}_{i,0}, \mathbf{u}_{i,1}, \cdots, \mathbf{u}_{i,H - 1} \right)$ denote the candidate state and input trajectories of turbine~$i$. The term $\Delta\mathbf{u}_{i,k}$ symbolizes the input vector of turbine~$i$ at time-step number $k$ measured relative to the reference $\left[ \frac{1}{3}~0\,\si{deg} \right]^\mathrm{T}$. This reference is used so that only deviation of the axial induction factor relative to its optimal value of $\frac{1}{3}$ is penalized in the cost function. $\mathbf{Q}$ is a weighting matrix that regulates the significance of this penalty. The set $\mathcal{N}_i$ contains the indices of all agents that are physically adjacent to turbine~$i$. Finally, $A_{\mathrm{OL}, i \rightarrow j}(\cdot)$ is a function that computes the overlap area between the rotors of turbines~$i$ and~$j$ and then normalizes this value based on the rotor swept area. 

Constraint~(\ref{Constraint - Initial conditions}) simply states that the initial condition along the prediction horizon must equate to the latest state measurement. Constraint~(\ref{Constraint - Dynamic model}) relates the candidate state and input vector trajectories via a model of the dynamics of turbine~$i$. This dynamic model $\mathbf{f}_i(\cdot)$ is estimated using feed-forward neural networks in Section~\ref{Subsection - Neural network}. Finally, constraints~(\ref{Constraint - Axial induction factor}) and~(\ref{Constraint - Yaw angle}) limit variations of the axial induction factor and yaw angle of turbine~$i$. In the case of yaw angles, values are limited to $\pm$10\,deg for operational safety. In the current work, the axial induction factors of all turbines are held fixed at the optimal operating point of $\frac{1}{3}$. In future works that attempt to directly maximize power production, the axial induction factor may be permitted to vary since its impact is considered in the power estimation function.

Constraint~(\ref{Constraint - Terminal set-point}) is the terminal set-point. It requires that the optimal state trajectory $\overline{\mathbf{x}}_i$ ends at a stationary state vector $\mathbf{x}_{i,s}$, which is a candidate steady-state. The computation of $\mathbf{x}_{i,s}$ requires solving the following optimization problem at each time-step:
\begin{equation} \label{Equation - Stationary optimization problem}
\min_{\overline{\mathbf{x}}_i,\mathbf{x}_{i,s},\overline{\mathbf{u}}_i,\mathbf{u}_{i,s}} \Delta\mathbf{u}_{i,s}^\mathrm{T} \mathbf{Q} \Delta\mathbf{u}_{i,s} + \sum_{j \in \mathcal{N}_i} \left[\frac{1}{\left| \mathcal{N}_i \right|} A_{\mathrm{OL}, i \rightarrow j}(\mathbf{x}_{i,s},\mathbf{x}_{j,s}) + \Delta\mathbf{u}_{j,s}^\mathrm{T} \mathbf{Q} \Delta\mathbf{u}_{j,s} \right],
\end{equation}
subject to constraints in Eq.~(\ref{Equation - Dynamic constraints}) along with
\begin{subequations}
	\begin{eqnarray}
	\mathbf{x}_{i,s} & = & \mathbf{f}_i(\mathbf{x}_{i,s},\mathbf{u}_{i,s}), \label{Constraint - Dynamic model, stationary}\\
	a_{i,s} & = & \frac{1}{3}, \label{Constraint - Axial induction factor, stationary}\\
	\left| \gamma_{i,s} \right| & \leq & 10\,\si{deg}, \label{Constraint - Yaw angle, stationary}
	\end{eqnarray}
\end{subequations}
where $\mathbf{u}_{i,s}$ is the candidate optimal stationary input vector of turbine~$i$, and $\Delta\mathbf{u}_{i,s}$ denotes $\mathbf{u}_{i,s}$ measured relative to the reference $\left[ \frac{1}{3}~0\,\si{deg} \right]^\mathrm{T}$. Constraint~(\ref{Constraint - Dynamic model, stationary}) requires that $\mathbf{x}_{i,s}$ and $\mathbf{u}_{i,s}$ correspond to a stationary point. Similar to Problem~(\ref{Equation - Dynamic optimization problem}), constraints~(\ref{Constraint - Axial induction factor, stationary}) and~(\ref{Constraint - Yaw angle, stationary}) serve as limits on variations of the axial induction factor and yaw angle at steady-state.

The difference between Problems~(\ref{Equation - Dynamic optimization problem}) and~(\ref{Equation - Stationary optimization problem}) is that the latter does not minimize the stage cost function throughout the prediction horizon. Instead, it only minimizes overlap areas and actuator deviation at steady-state. It therefore computes an optimal stationary point independent of the trajectory that leads to this point. However, since the constraints require that $\mathbf{x}_{i,H} = \mathbf{x}_{i,s}$, the computed optimal steady-state is guaranteed to be reachable given the current state measurement. The vector $\mathbf{x}_{i,s}$ is therefore a candidate \textit{reachable} steady-state.

\subsection{Feed-forward artificial neural network} \label{Subsection - Neural network}

In order to hasten the dynamic optimization process of the DEMPC algorithm, feed-forward neural networks are used to approximate the dynamic model $\mathbf{f}_i(\cdot)$ for any turbine~$i$. This estimated model is solely a function of the states and inputs of turbine~$i$ and thus treats aerodynamic coupling as disturbance or model uncertainty.

\begin{figure}
	\centering
	\includegraphics[width=4in]{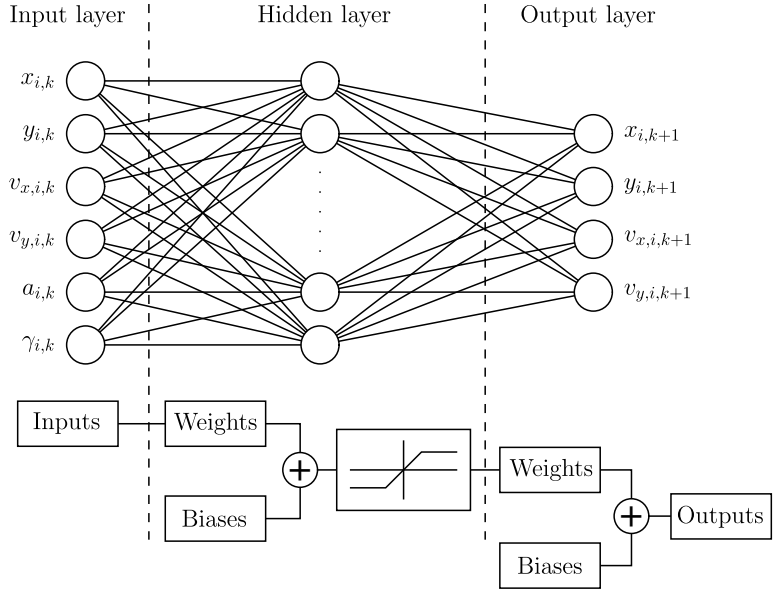}
	\caption{Schematic of the local neural network structure used to identify the dynamic model of each floating wind turbine.} \label{Figure - Neural network}
\end{figure}

Each neural network takes the form shown in Fig.~\ref{Figure - Neural network}. The six input neurons represent the states $\mathbf{x}_{i,k}$ and inputs $\mathbf{u}_{i,k}$ of the corresponding turbine at some time-step number $k$ along the prediction horizon. This input layer feeds into a single hidden layer consisting of 20 neurons. The output layer contains four neurons representing the turbine states $\mathbf{x}_{i,k + 1}$ at the subsequent time-step number $k + 1$.

To generate training data for tuning the neural networks, FOWFSim-Dyn simulations consisting of 10$^5$ time-steps with sampling periods of 60\,sec were completed for each wind farm configuration that was examined. The free stream wind velocity was fixed at 8\,m/s in the direction of the positive $\hat{x}$ axis. The initial conditions of each simulation were computed by operating all turbines with an axial induction factor of $\frac{1}{3}$ and a yaw angle of 0\,deg. Then, over the course of the simulations, turbine inputs were randomly varied with uniform probability of 0.1 (\textit{i.e.} there was a 10\,\% chance that the input values would change at each new time-step). The logic behind this variation approach was to allow enough time for the turbine platforms to respond to each change in input values. Axial induction factors were varied between 0.2~and 0.4, while yaw angles were varied between $-$20~and $+$20\,deg.

\begin{figure}
	\centering
	\includegraphics[width=4in]{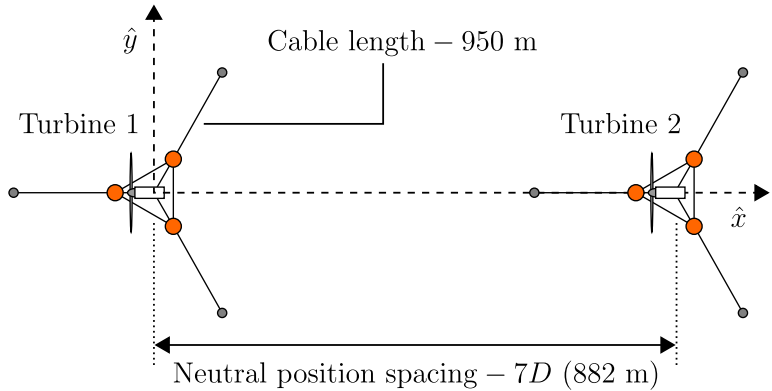}
	\caption{Schematic of a two-turbine floating offshore wind farm simulated in the current work. Wind turbine and floating platform properties are based on the National Renewable Energy Laboratory's (NREL's) baseline designs presented by Jonkman \textit{et al.}~\cite{Jonkman2009} and Robertson \textit{et al.}~\cite{Robertson2014}. The sole modification is that mooring line lengths have been increased from $835$~to $950\,\si{m}$ to permit greater platform displacement.} \label{Figure - 2 turbine farm}
\end{figure}

Sample validation data is provided for the two-turbine floating wind farm described in Fig.~\ref{Figure - 2 turbine farm}. The turbine and platform designs are detailed in the works of Jonkman \textit{et al.}~\cite{Jonkman2009} and Robertson \textit{et al.}~\cite{Robertson2014}, except mooring line lengths have been increased from 835~to 950\,m to increase platform range of motion. In their neutral positions, the platforms are spaced seven rotor diameters apart, which equates to 882\,m.

For validation, FOWFSim-Dyn and the neural networks were given the same randomly generated initial condition and turbine input sequences over 60~time-steps (\textit{i.e.} 3,600\,sec). Their predictions of the system response are compared for both turbines~1 and~2 in Fig.~\ref{Figure - NN validation, Turbine 1} and Fig.~\ref{Figure - NN validation, Turbine 2}, respectively. These plots correspond to a single run, and have been provided for qualitative inspection. It is apparent that neural network predictions for turbine~1 more closely match simulation results. This outcome is expected since turbine~1 is the leading machine and is thus uninfluenced by aerodynamic coupling. The behaviour of turbine~2 is affected by the states and inputs of turbine~1 however, which induce greater prediction errors in the behavior of turbine~2 since its neural network does not consider dynamic coupling.

\begin{figure}
	\centering
	\includegraphics[width=4in]{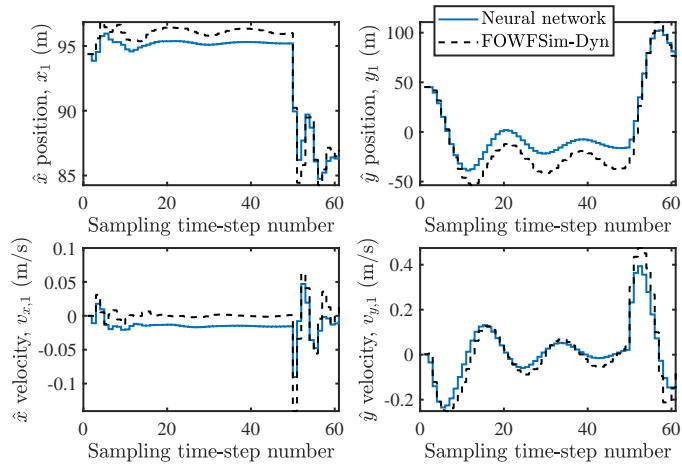}
	\caption{Sample validation data for the neural network of  the upstream machine (\textit{i.e.} turbine~1) in a two-turbine floating wind farm.} \label{Figure - NN validation, Turbine 1}
\end{figure}

\begin{figure}
	\centering
	\includegraphics[width=4in]{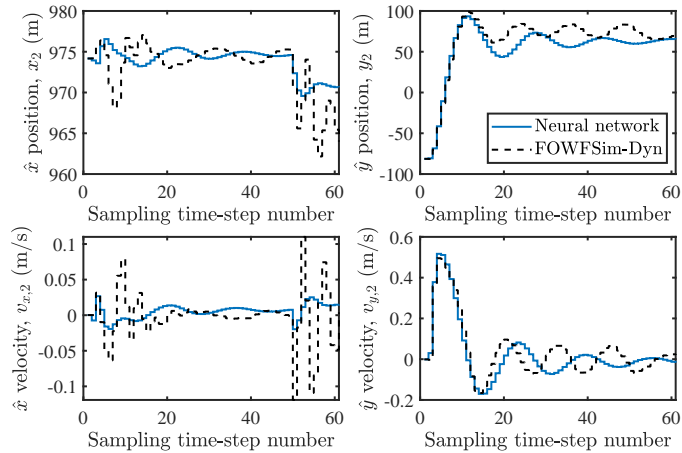}
	\caption{Sample validation data for the neural network of the upstream machine (\textit{i.e.} turbine~2) in a two-turbine floating wind farm.} \label{Figure - NN validation, Turbine 2}
\end{figure}

For a quantitative assessment of neural network performance, ten different runs were completed to acquire the average root-mean-square-error (RMSE) of each neural network output. This information is listed in Table~\ref{Table - RMSE values}. Neural network predictions of $x_i$ deviate from FOWFSim-Dyn results with RMSE values of $0.94$~and $5.79\,\si{m}$ for turbines~1 and~2, respectively. The greater RMSE associated with turbine~2 is attributed to the absence of dynamic coupling in the neural networks. This trend reverses in the case of $y_i$ predictions. Since turbine~1 is unaffected by any upstream wakes, it encounters higher wind speeds and thus displaces a greater distance in the crosswind direction relative to turbine~2 when operating with yaw misalignment. As a result, the RMSE value of $y_1$ is inflated in comparison to that of $y_2$. These arguments apply equally well with regards to platform velocity RMSE values. Successful control of the current system therefore requires that local EMPC algorithms are capable of rejecting these prediction errors. A description of the social hierarchy-based DEMPC scheme now follows.

\begin{table}
	\centering
	\caption{List of RMSE values between neural network predictions and FOWFSim-Dyn predictions.}
	\label{Table - RMSE values}
	\scriptsize
	\begin{tabular}{|c|c|c|c|c|}
		\hline
		Turbine & $x_i\,(\si{m})$ & $y_i\,(\si{m})$ & $v_{x,i}\,(\si{m/s})$ & $v_{y,i}\,(\si{m/s})$ \\ \hline \hline
		1 & 0.94       & 14.68      & 0.02        & 0.08        \\ \hline
		2 & 5.79       & 11.89      & 0.05        & 0.06        \\ \hline
	\end{tabular}
\end{table}

\subsection{Social hierarchy-based DEMPC algorithm}

If multiple optimal solutions exist in an optimal control problem (\textit{i.e.} the cost functions are non-convex), and decisions are made in a synchronous and distributed manner by multiple agents, then there exists a potential for conflicting decisions~\cite{Kheirabadi2020a}. It is due to this logic that convergence of the decisions made by multiple agents cannot be guaranteed in parallel distributed optimal control problems~\cite{Liu2012a}. The social hierarchy-based DEMPC algorithm used in this work was therefore originally developed to address non-convexity in distributed optimal control problems. The essential function of this algorithm is that all agents self-organize in a fully distributed manner into a social hierarchy that dictates the order in which each agent makes a decision.

The algorithm borrows from evolutionary principles. Namely, if an agent's current level along a social hierarchy yields decisions that conflict with those of its neighbors, then the agent's current level is detrimental to performance and must be randomly varied. The method with which conflicting decisions are identified is now described. At any iteration, each agent possesses assumed values of the state and input trajectories of its neighbors. Based on these assumptions, and given some existing social hierarchy that determines the order in which agents make decisions, the agent updates its own optimal trajectories and calculates the resulting value of its neighborhood cost function. We refer to this value as the \textit{naive} cost function. After receiving updated optimal state and input trajectories from its neighbors, the agent recomputes the value its neighborhood cost function. We refer to this updated value as the \textit{informed} cost function. If the informed cost function is improved in comparison to the naive cost function, then the agent's decisions are mutually beneficial with those of its neighbors. Otherwise, the decisions are conflicting and the agent must randomly change its place along the social hierarchy. It was proved in our previous work that, for any interaction topology, at least one social hierarchy exists that enables the agents to avoid conflicting decisions; thus establishing a guarantee of convergence.

Another unique property of this algorithm is that it was developed without the presumption of terminal set-points that are known \textit{a priori}. Instead, it was intended for the agents to determine these terminal targets in a fully distributed manner. With mild similarity to the approach taken by Kohler \text{et al.}~\cite{Kohler2018b}, the agents compute their terminal set-points at each time-step by first solving an optimization problem which minimizes their cost functions at steady-state. This steady-state set-point is utilized as a terminal constraint in the subsequent optimization problem that yields optimal dynamic state and input trajectories. The computation of the steady-state set-points also makes use of the social hierarchy-based approach to eliminate conflicting decisions.

\subsection{Effects of dynamic coupling} \label{Subsection - Effects of dynamic coupling}

In the current work, feed-forward neural networks are used to locally estimate the dynamic model of each floating wind turbine. These estimated models are better suited for dynamic optimization due to their low computation time requirements. Since these neural networks are tuned locally, they are solely functions of the inputs and states of their corresponding turbines. They therefore do not consider the inputs and states of turbines with which aerodynamic coupling exists. Dynamic coupling may thus be interpreted as a source of disturbance or model uncertainty in our work.

\begin{figure}
	\centering
	\includegraphics[width=3in]{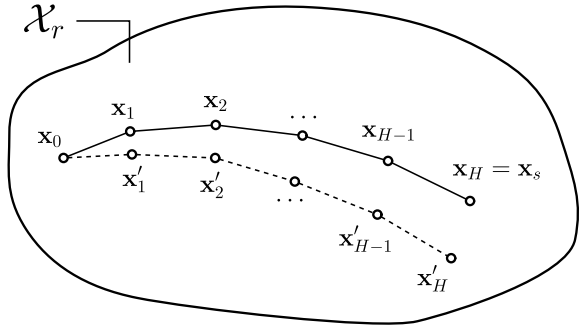}
	\caption{Schematic showing the impacts of disturbance/model uncertainty on terminal equality constraint satisfaction in EMPC} \label{Figure - EMPC terminal dist. effects}
\end{figure}

EMPC is inherently capable of rejecting disturbance or model uncertainty that is appropriately bounded in magnitude given the dynamics of the plant. The impact of disturbances in EMPC is first demonstrated in the context of terminal equality constraint satisfaction in Fig.~\ref{Figure - EMPC terminal dist. effects}. Let the set $\mathcal{X}_r$ contain all initial state vectors $\mathbf{x}_0$ from which a terminal set-point $\mathbf{x}_s$ is reachable in $H$ time-steps while satisfying process constraints. As long as $\mathbf{x}_0$ is contained within the set $\mathcal{X}_r$, then an input trajectory $\overline{\mathbf{u}} = \left( \mathbf{u}_0, \mathbf{u}_1, \cdots, \mathbf{u}_{H - 1} \right)$ exists that generates a state trajectory $\overline{\mathbf{x}} = \left( \mathbf{x}_0, \mathbf{x}_1, \cdots, \mathbf{x}_H \right)$ which ends at $\mathbf{x}_H = \mathbf{x}_s$. Disturbance or model uncertainty causes a deflection of the state trajectory from $\overline{\mathbf{x}}$ to $\overline{\mathbf{x}}' = \left( \mathbf{x}_0, \mathbf{x}_1', \cdots, \mathbf{x}_H' \right)$ as the system moves forward in time. Provided that the magnitude of disturbance or model uncertainty is appropriately bounded, then $\mathbf{x}_H'$ will not leave the set $\mathcal{X}_r$. As a result, even if the system follows the deflected path from $\mathbf{x}_0$ to $\mathbf{x}_H'$, then a path from $\mathbf{x}_H'$ to $\mathbf{x}_s$ is guaranteed to exist. If, on the other hand, the magnitude of disturbance is large enough to cause $\mathbf{x}_H'$ to leave $\mathcal{X}_r$, the satisfaction of the terminal constraint cannot be guaranteed and system redesign is recommended.

Next, the impact of disturbance on the satisfaction of trajectory constraints is discussed. Let the set $\mathcal{X}$ contain all acceptable values of the state vector $\mathbf{x}_k$ for all $k \in \mathbb{I}_{0:H}$. In other words, the state trajectory $\overline{\mathbf{x}}$ must always be contained within $\mathcal{X}$. As before, disturbance or model uncertainty deflects $\overline{\mathbf{x}}$ to $\overline{\mathbf{x}}'$ as the system moves forward in time. Provided that the magnitude of disturbance is appropriately bounded, then $\overline{\mathbf{x}}'$ will not leave the set $\mathcal{X}$ and state constraints will be satisfied. This expectation is conservative however. If the magnitude of disturbance is not appropriately bounded, then the technique of constraint tightening~\cite{Chisci2001} may be used to ensure constraint satisfaction.

\begin{figure}
	\centering
	\includegraphics[width=3in]{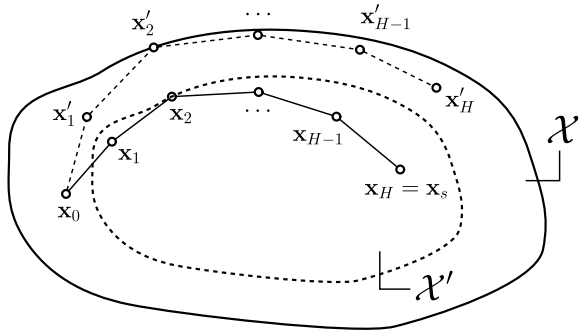}
	\caption{Schematic demonstrating the method of constraint tightening for process constraint satisfaction in EMPC.} \label{Figure - EMPC process dist. effects}
\end{figure}

The mechanism of constraint tightening is demonstrated in Fig.~\ref{Figure - EMPC process dist. effects}. Once again, it is required that $\overline{\mathbf{x}}$ remains within the set $\mathcal{X}$. To ensure this outcome, the set $\mathcal{X}$ may be duplicated and \textit{tightened} to form the set $\mathcal{X}'$. The dynamic optimization problem must then be required to maintain $\overline{\mathbf{x}}$ within $\mathcal{X}'$. If $\mathcal{X}'$ is sized accordingly, then, even with the worst-case disturbance scenario, the deflected state trajectory $\overline{\mathbf{x}}'$ is guaranteed to remain within the original process constraint set $\mathcal{X}$.

\section{Results and discussion} \label{Section - Results and discussion}

The current section presents simulation results corresponding to the implementation of the DEMPC scheme outlined in Section~\ref{Section - Controller design} on floating wind farms of different sizes. Each simulated wind farm possesses a row configuration similar to that shown in Fig.~\ref{Figure - 2 turbine farm}. Inter-turbine neutral position spacings are seven rotor diameters and mooring line cable lengths are 950\,m, while the sizes of the wind farms range from $1 \times 2$ to $1 \times 5$. First, the impact of wind speed and direction variation on controller performance is assessed for a $1 \times 2$ wind farm. Then, the performance and behaviour of larger array sizes are investigated.

\subsection{Wind velocity variation}

The neural networks described in Section~\ref{Subsection - Neural network} were tuned using data corresponding to 8\,m/s free stream wind aligned with the positive $\hat{x}$ axis. As the free stream wind speed and direction deviate from this condition, the error in neural network outputs will increase and cause detriment to controller performance. We therefore assess in this subsection the impacts of free stream wind velocity fluctuations on platform motion and optimal power production.

\begin{figure}
	\centering
	\includegraphics[width=4in]{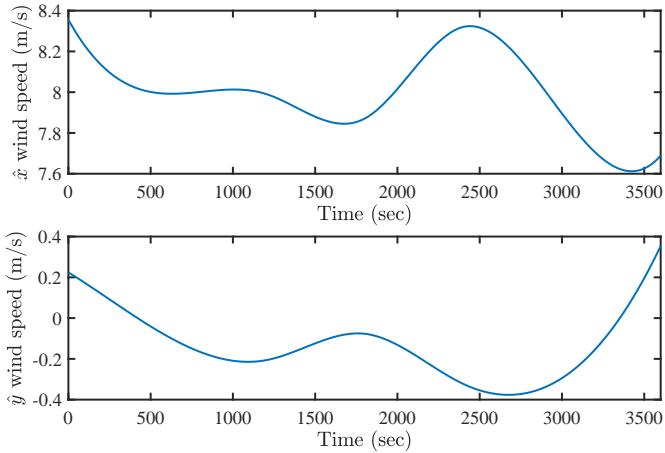}
	\caption{Sample randomly generated free stream wind speed evolutions in the $\hat{x}$ and $\hat{y}$ directions corresponding to a perturbation parameter of $\sigma_\infty$ = 5\,\%.} \label{Figure - Sample wind velocity}
\end{figure}

To quantify velocity fluctuations, we define the parameter $0 \leq \sigma_\infty \leq 1$ and use it to randomly perturb, with uniform probability, the ten-minute averaged free stream wind velocity vector. Therefore, at some ten-minute mark $t$, the free stream wind velocity vector $\mathbf{V}_\infty(t)$ is computed as follows: 
\begin{equation}
\mathbf{V}_\infty(t) = \overline{\mathbf{V}}_\infty +
\begin{bmatrix}
\mathrm{rand}_{\pm \sigma_\infty \left\Vert \overline{\mathbf{V}}_\infty \right\Vert}\\
\mathrm{rand}_{\pm \sigma_\infty \left\Vert \overline{\mathbf{V}}_\infty \right\Vert}
\end{bmatrix},
\end{equation}
where $\overline{\mathbf{V}}_\infty$ is the baseline free stream velocity vector of 8\,m/s aligned with the positive $\hat{x}$ axis, and $\mathrm{rand}_{\pm \sigma_\infty \left\Vert \overline{\mathbf{V}}_\infty \right\Vert}$ is a function that generates a random number between $-\sigma_\infty \left\Vert \overline{\mathbf{V}}_\infty \right\Vert$ and $+\sigma_\infty \left\Vert \overline{\mathbf{V}}_\infty \right\Vert$. Spline interpolation is then used to generate smoother curves with 0.1~sec resolution using these randomly generated ten-minute averages. Sample free stream wind velocity curves in the $\hat{x}$ and $\hat{y}$ directions corresponding to $\sigma_\infty =$ 5\,\% are plotted in Fig.~\ref{Figure - Sample wind velocity}.

\subsection{Social hierarchy-based DEMPC properties}

The relevant properties of the social hierarchy-based DEMPC algorithm are listed in Table~\ref{Table - DEMPC properties}. Only two social hierarchy levels are used since, in order to prevent conflicting decisions, each turbine must be given the opportunity to compute its optimal trajectory out of synchrony with its adjacent neighbors. Within each sampling period, each turbine is granted three iterations to compute its optimal stationary point by solving Problem~(\ref{Equation - Stationary optimization problem}), as well as three additional iterations to compute its optimal state and input trajectories by solving Problem~(\ref{Equation - Dynamic optimization problem}).

\begin{table}
	\centering
	\caption{List of properties relevant to the social hierarchy-based DEMPC algorithm.}
	\label{Table - DEMPC properties}
	\scriptsize
	\begin{tabular}{|l|c|}
		\hline
		Property                            & Value \\ \hline \hline
		Number of social hierarchy levels            & 2              \\ \hline
		Number of iterations per sampling period     & 3              \\ \hline
		Number of prediction horizon time-steps, $H$ & 5              \\ \hline
		Input weight matrix, $\mathbf{Q}$            & $\begin{bmatrix} 1 & 0 \\ 0 & 1 \end{bmatrix}$           \\ \hline
	\end{tabular}
\end{table}

The length of the prediction horizon is set to $H = 5$. Since the sampling period of each controller is 60\,sec, this value of $H$ permits prediction of 5\,min into future. Finally, the input weight matrix $\mathbf{Q}$ used in Problems~(\ref{Equation - Stationary optimization problem}) and~(\ref{Equation - Dynamic optimization problem}) is set the identity matrix. Since the overlap areas between turbine rotors are normalized in the cost functions of Problems~(\ref{Equation - Dynamic optimization problem}) and~(\ref{Equation - Stationary optimization problem}), input weights of unity render the optimization problem approximately circular. An identity matrix was therefore used a starting point for $\mathbf{Q}$ and the corresponding results were deemed acceptable.

Regardless of the wind farm size, the time required by the DEMPC algorithm to compute optimal state and input trajectories \textit{per turbine} ranged from 10~to 50\,sec. This range of durations is smaller than the sampling period of 60\,sec; the social hierarchy-based DEMPC algorithm may therefore be implemented in real-time. It is important to note that, if a sequential DEMPC algorithm~\cite{Kuwata2007,Richards2007} had been used, then the total computation time would be a summation of the durations required by all turbines. This property exists since a sequential algorithm requires that individual turbines take turns to make decisions. In other words, even for a two-turbine wind farm, the total required computation time would range from 20~to 100\,sec, which exceeds the 60\,sec sampling period.

\subsection{Effects of wind velocity variations}

The impact of time-varying free stream wind velocity on controller performance is assessed for a $1 \times 2$ wind farm. Evolutions of turbine $\hat{y}$ positions for different $\sigma_\infty$ values are shown in Fig.~\ref{Figure - Y-pos traj, 2 turbines}. As desired, the two floating wind turbines relocate in opposite directions in all cases to minimize their respective rotor overlap areas. Specifically, the turbines shift by approximately one rotor radius (\textit{i.e.} 60\,m), which results in a total clearance of one rotor diameter between their nacelle centerlines.

\begin{figure}
	\centering
	\includegraphics[width=4.5in]{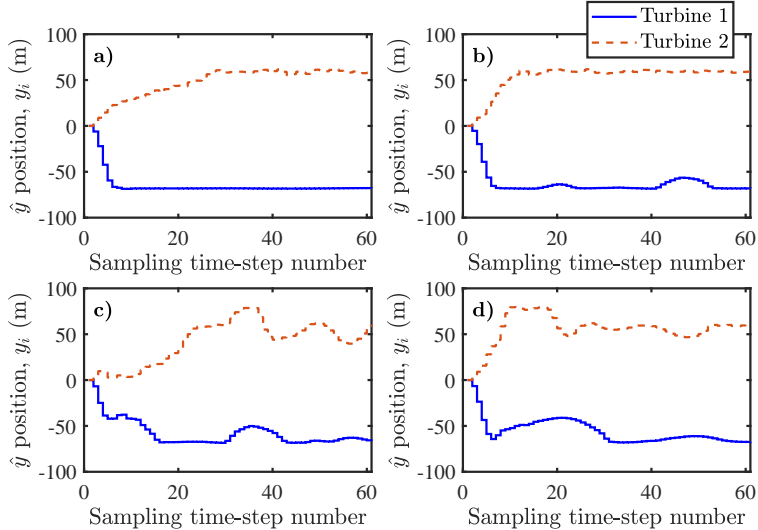}
	\caption{Evolution of $\hat{y}$ positions in a two-turbine floating wind farm that is optimally controlled by the DEMPC from Section~\ref{Section - Controller design}. The four plots correspond to different wind velocity perturbations of a) $\sigma_\infty =$ 5\,\%, b) $\sigma_\infty =$ 10\,\%, c) $\sigma_\infty =$ 15\,\%, and d) $\sigma_\infty =$ 20\,\%.} \label{Figure - Y-pos traj, 2 turbines}
\end{figure}

The impact of neural network prediction error is also evident in two aspects. First, regardless of the wind velocity perturbation parameter $\sigma_\infty$, the downstream turbine's $\hat{y}$ position shows greater variability over time. The reason is that the dynamics of the downstream turbine (\textit{i.e.} turbine~2) are influenced by the states and inputs of the upstream machine. Since this coupling was not explicitly considered in the machine learning process, the neural network of turbine~2 is subject to greater prediction error. This error translates to deflections of optimal state trajectories as discussed in Section~\ref{Subsection - Effects of dynamic coupling}.

Second, as the wind velocity perturbation $\sigma_\infty$ is increased from 5~to 20\,\%, the $\hat{y}$ positions of both turbines show greater variability over time. This outcome is also explained by neural network prediction error. The neural networks have been tuned using data corresponding to a free stream wind velocity of 8\,m/s aligned with the $\hat{x}$ axis. As the wind velocity deviates from this reference, the neural network uncertainties are exacerbated, once again leading to deflections in state trajectories relative to computes optimal values.

Evolutions of wind farm power production for different $\sigma_\infty$ values are shown in Fig.~\ref{Figure - Farm power traj, 2 turbines}. This figure displays power production trends corresponding to both DEMPC and \textit{greedy}\footnote{Greedy operation implies the absence of wind farm control. That is to say, each wind turbine is operated with an axial induction factor and yaw angle that maximize its own power output; these greedy values are $a_i = \frac{1}{3}$ and $\gamma_i =$ 0\,deg.} operation subject to identical free stream wind conditions. The greedy performance curves demonstrate the amount of power produced without wind farm control. Any rise in power production in the greedy case is consequently purely a result of free stream wind speed fluctuations. The benefit incurred by implementing YITuR is therefore quantified by the difference between the DEMPC and greedy power trajectories. With 5\,\% variability in the wind velocity, YITuR yields an 18.4\,\% gain in energy production relative to greedy operation throughout the hour-long simulation.

\begin{figure}
	\centering
	\includegraphics[width=4.5in]{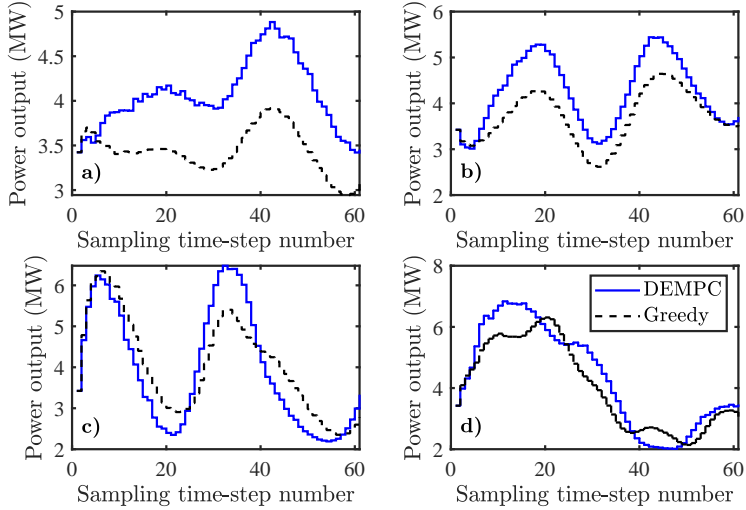}
	\caption{Evolution of power production in a two-turbine floating wind farm that is optimally controlled by the DEMPC from Section~\ref{Section - Controller design}. For comparison, power production under greedy operation is also demonstrated. The four plots correspond to different wind velocity perturbations of a) $\sigma_\infty =$ 5\,\%, b) $\sigma_\infty =$ 10\,\%, c) $\sigma_\infty =$ 15\,\%, and d) $\sigma_\infty =$ 20\,\%.} \label{Figure - Farm power traj, 2 turbines}
\end{figure}

An additional observation from Fig.~\ref{Figure - Farm power traj, 2 turbines} is that the increased variability of the free stream wind velocity diminishes wind farm controller performance. As $\sigma_\infty$ is raised from 5~to 20\,\%, the relative gain in wind farm energy production obtained by switching from greedy operation to optimal control decreases from 18.4~to 7.3\,\%. This outcome is partly caused by the aforementioned rise in neural network prediction error that results from wind speed and direction deviations from the values used for neural network tuning. As seen in Fig.~\ref{Figure - Y-pos traj, 2 turbines}, this error causes turbine $\hat{y}$ displacements to decrease for brief periods, which increases rotor overlap areas. Another factor is that, as the free stream wind direction varies and misaligns from the row of turbines, wake overlap naturally subsides. As a result, wind farm control fails to yield much benefit relative to greedy operation.

Platform displacements in the $\hat{x}$ direction are not considered in the DEMPC optimization problems; their progressions are therefore entirely dictated by the free stream wind velocity and aerodynamic coupling. Insight may nonetheless be gained by examining their trends. Evolutions of turbine $\hat{x}$ displacements (\textit{i.e.} relative to neutral positions) are shown in Fig.~\ref{Figure - X-pos traj, 2 turbines}. The first observation is that turbine~2 consistently displaces a smaller distance in the downwind direction in comparison to turbine~1 for all wind velocity scenarios. The reason is simply that, as the downstream machine, turbine~2 encounters slower wind speeds on average due to the presence of the wake produced by its upstream neighbor.

\begin{figure}
	\centering
	\includegraphics[width=4.5in]{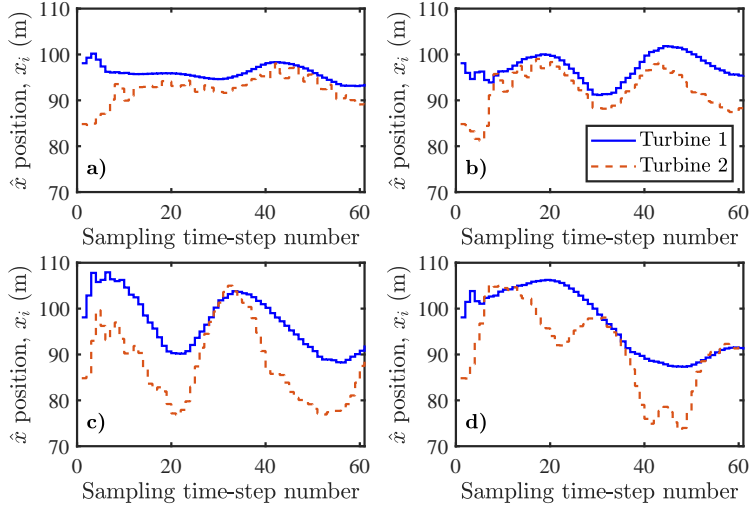}
	\caption{Evolution of $\hat{x}$ positions in a two-turbine floating wind farm that is optimally controlled by the DEMPC from Section~\ref{Section - Controller design}. The four plots correspond to different wind velocity perturbations of a) $\sigma_\infty =$ 5\,\%, b) $\sigma_\infty =$ 10\,\%, c) $\sigma_\infty =$ 15\,\%, and d) $\sigma_\infty =$ 20\,\%.} \label{Figure - X-pos traj, 2 turbines}
\end{figure}

A second trend from Fig.~\ref{Figure - X-pos traj, 2 turbines} shows that the downwind displacement of turbine~2 is more volatile over time. This behaviour is once again the result of aerodynamic coupling. As turbine~1 adjusts its yaw angle and relocates its platform, the resultant effects on the wind field are transported downwind to the location of turbine~2. The downstream machine therefore encounters greater variability in the effective wind velocity that is incident upon its rotor. Furthermore, larger variations in the free stream wind velocity cause greater volatility in the downwind displacement of both turbines. In the case of $\sigma_\infty =$ 5\,\%, the $\hat{x}$ displacement of turbine~2 ranges from 85~to 100\,m. This range expands to 75~to 105\,m at $\sigma_\infty =$ 20\,\%. This outcome indicates that some additional control function may be required to limit excessive platform displacements in the downwind direction. This function may take the form of constraints in the DEMPC optimization process or, alternatively, downwind position regulation may be achieved via wind turbine-level control systems.

\subsection{Performance of different wind farm sizes}

In this subsection, the impact of increasing the wind farm array size is assessed while maintaining $\sigma_\infty =$ 5\,\%. Evolutions of wind farm power production for wind farm sizes ranging from $1 \times 2$ to $1 \times 5$ are shown in Fig.~\ref{Figure - Farm power traj, 2 turbines} for both optimal operation using DEMPC and greedy operation. In all cases, the energy produced by the wind farm over the simulated hour increases by approximately 20\,\% when optimal control is used in place of greedy operation. This result contradicts the findings of our stationary optimization study~\cite{Kheirabadi2020}, wherein the relative gain in wind farm efficiency increased from 21.8~to 42.7\,\% as the wind farm size was varied from $2 \times 2$ to $7 \times 7$.

\begin{figure}
	\centering
	\includegraphics[width=4.5in]{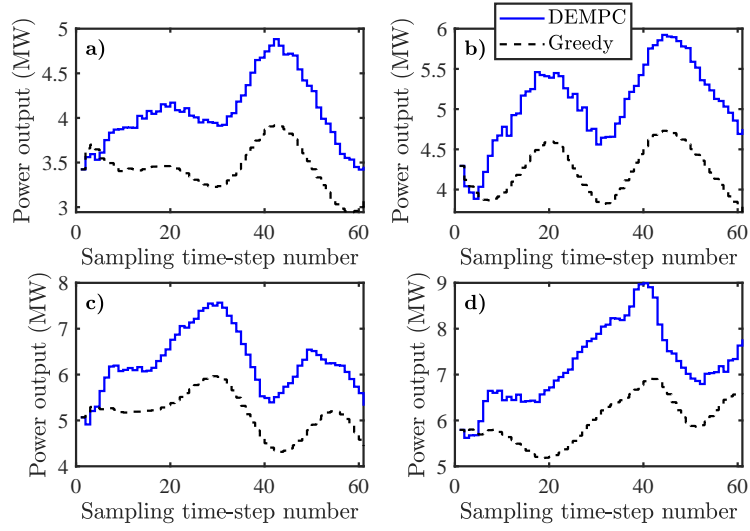}
	\caption{Evolution of power production in floating wind farms of different sizes that are optimally controlled by the DEMPC from Section~\ref{Section - Controller design}. For comparison, power production under greedy operation is also demonstrated. The wind velocity perturbation is fixed at $\sigma_\infty =$ 5\,\%. The four plots correspond to different wind farm configurations of a) $1 \times 2$, b) $1 \times 3$, c) $1 \times 4$, and d) $1 \times 5$.} \label{Figure - Farm power traj, multiple turbines}
\end{figure}

One possible explanation for this outcome is that, as the wind farm size increases, the trailing machines are subject to dynamic coupling from a larger number of upstream counterparts. As a result, their neural networks are subject to greater prediction error, which renders their optimal controllers less effective. This explanation is validated by plotting evolutions of turbine $\hat{y}$ positions for the different wind farm sizes in Fig.~\ref{Figure - Y-pos traj, multiple turbines}. It is evident that, although $\sigma_\infty$ remains unchanged, the $\hat{y}$ position of each subsequent downstream turbine displays greater variability in comparison to those of its upstream neighbors. The $\hat{y}$ position trajectory of turbine~5 in Fig.~\ref{Figure - Y-pos traj, multiple turbines}d defies this trend; however, the $\hat{y}$ position trajectory of turbine~3 is consistently more volatile than those of turbines~1 and~2 in all relevant plots, and the $\hat{y}$ position trajectory of turbine~4 is consistently more volatile that that of turbine~3 in all relevant plots. This increased volatility results in longer moments of increased overlap between turbine rotors.

\begin{figure}
	\centering
	\includegraphics[width=4.5in]{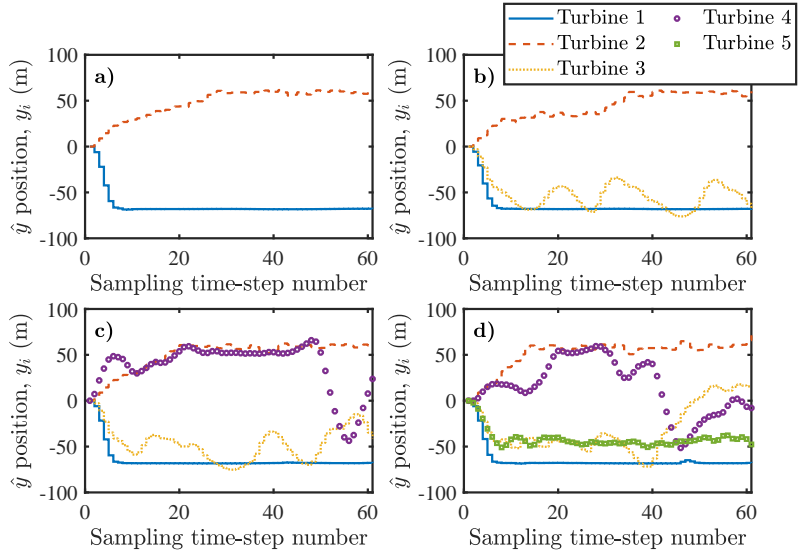}
	\caption{Evolution of $\hat{y}$ positions in floating wind farms of different sizes that are optimally controlled by the DEMPC from Section~\ref{Section - Controller design}. The wind velocity perturbation is fixed at $\sigma_\infty =$ 5\,\%. The four plots correspond to different wind farm configurations of a) $1 \times 2$, b) $1 \times 3$, c) $1 \times 4$, and d) $1 \times 5$.} \label{Figure - Y-pos traj, multiple turbines}
\end{figure}

A final observation from Fig.~\ref{Figure - Y-pos traj, multiple turbines} that is worth point out is the fact that the social hierarchy-based DEMPC algorithm successfully prevents conflicting decisions in all simulated cases. For each wind farm size, neighboring floating wind turbines are relocated in opposite directions, which is the most effective modified layout for minimizing adjacent overlap areas.

Evolutions of turbine $\hat{x}$ positions for the different wind farm sizes are plotted in Fig.~\ref{Figure - X-pos traj, multiple turbines}. As observed from $\hat{x}$ displacements in the case of the two-turbine wind farm from Fig.~\ref{Figure - X-pos traj, 2 turbines}, downstream machines shift a smaller distance on average in the downwind direction due to reduced incident wind speeds while also experiencing greater volatility in $\hat{x}$ displacements over time due to aerodynamic coupling. However, it is observed in Fig.~\ref{Figure - X-pos traj, multiple turbines} that, with each subsequent downstream machine, additional downwind displacements gradually diminish. For instance, over the course of the 3,600\,sec simulation from Fig.~\ref{Figure - X-pos traj, multiple turbines}d, the average downwind displacement of turbine~1 is the largest at 96.1\,m. This value drops to 90.9~and 86.0\,m for turbines~2 and~3, and then to 82.8\,m for both turbines~4 and~5. Clearly, the wind velocity deficit resulting from the compounding of overlapping wakes ultimately converges to a finite value. This result is significant in the context of turbine repositioning. It indicates that, regardless of the size of the wind farm, turbines that are located far downstream may still encounter large enough wind speeds to generate sufficient platform displacements for rotor overlap reduction.

\begin{figure}
	\centering
	\includegraphics[width=4.5in]{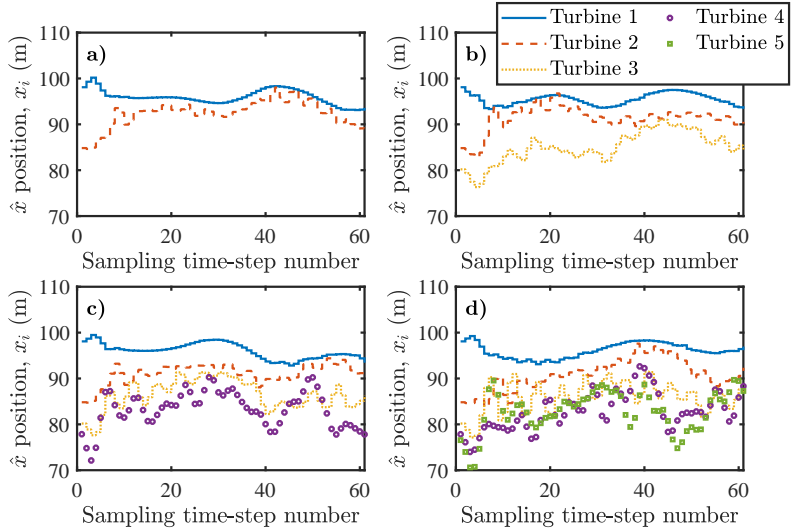}
	\caption{Evolution of $\hat{x}$ positions in floating wind farms of different sizes that are optimally controlled by the DEMPC from Section~\ref{Section - Controller design}. The wind velocity perturbation is fixed at $\sigma_\infty =$ 5\,\%. The four plots correspond to different wind farm configurations of a) $1 \times 2$, b) $1 \times 3$, c) $1 \times 4$, and d) $1 \times 5$.} \label{Figure - X-pos traj, multiple turbines}
\end{figure}

\section{Conclusion} \label{Section - Conclusions}

This paper presented the first real-time implementation of wind farm control for power maximization in floating offshore wind farms. The specific mechanism for power maximization involved relocating wind turbines using aerodynamic forces to minimize overlap areas between the rotors of adjacent turbines.

Floating wind farms were simulated using FOWFSim-Dyn~\cite{Kheirabadi2020b}, which is a simulation tool designed specifically for capturing floating platform motion coupled with wind farm aerodynamics. The wind farm controller was based on a social hierarchy-based distributed model predictive control (DEMPC) algorithm~\cite{Kheirabadi2020a} developed to address convergence challenges in non-convex distributed optimal control problems. Finally, feed-forward neural networks were used to model floating turbine dynamics during the dynamic optimization process of the DEMPC algorithm.

Conclusions and recommendations that have been derived from the results of this work are now discussed. Neural networks were effectively used to predict floating wind turbine behaviour for wind farm rows with layouts of $1 \times 3$ or smaller while neglecting dynamic coupling. Future work should establish error bounds on neural network predictions while taking into consideration the effects of dynamic coupling for improved performance in larger wind farms.

Deviations in the free stream wind velocity from the baseline value at which neural networks were tuned diminished controller performance. Future work may consider the free stream wind speed and direction as inputs to the neural networks or, alternatively, unique neural networks may be identified for multiple wind speed and direction ranges.

In larger wind farms, downstream machines demonstrated high variability in platform positions over time. This outcome resulted from increased neural network prediction error due to dynamic coupling. Once the bounds on these uncertainties have been established, future work may implement constraint tightening methods to limit the variability of floating platform positions.

The current controller maximized wind farm power production indirectly by minimizing the overlap areas between the rotors of adjacent floating turbines. Future work should use some machine learning scheme to identify power output functions for each turbine to serve as cost functions in DEMPC, thus permitting direct maximization of wind farm power production.

In the current work, a single row of wind turbines aligned with the predominant free stream wind direction was examined. Identifying which pairs of turbines encounter strong aerodynamic coupling was therefore a trivial task. In more complex wind farm configurations, such as gridded layouts, the pairings of turbines that interact aerodynamically is determined by the wind direction at any given instant in time. Future work may therefore extend the cost functions used by the DEMPC algorithm to capture the effects of wind direction on turbine pairings.

The current paper was concerned solely with power maximization as the wind farm control objective. Future work may assess alternative objectives. For instance, at sufficiently large wind speeds, the primary concern of a wind farm operator shifts from power maximization to load reduction~\cite{Knudsen2015}. This shift may be accounted for via adjustments to the cost functions used in optimal control algorithms. Specifically, cost functions may be defined such that the turbines track power set-points while minimizing blade and tower vibrations among their neighborhoods. The role that YITuR may play in such a control problem should be investigated.

The optimal state and input trajectories computed using DEMPC in the current work should not be implemented directly. Instead, they should serve as set-points for turbine-level controllers that consider additional dynamic phenomena such as platform oscillation. Future work should therefore integrate wind farm-level control with lower controllers capable of rejecting disturbances caused by waves and wind gusts with the aim of minimizing turbine loads and platform oscillation.

Finally, all state information was assumed to be available in the current work. Future work should examine sensing and observer techniques for estimating floating platform positions and velocities.

\section*{Acknowledgment}

The authors are grateful for the financial support provided by the Natural Sciences and Engineering Research Council of Canada (NSERC).

\bibliographystyle{unsrt}
\bibliography{Library.bib}

\begin{thebibliography}{10}

\bibitem{Nilsson2015}
Karl Nilsson, Stefan Ivanell, Kurt~S. Hansen, Robert Mikkelsen, Jens~N.
  S{\o}rensen, Simon-Philippe Breton, and Dan Henningson.
\newblock {Large-eddy simulations of the Lillgrund wind farm}.
\newblock {\em Wind Energy}, 18(3):449--467, 2015.

\bibitem{Kheirabadi2019a}
Ali~C. Kheirabadi and Ryozo Nagamune.
\newblock {A quantitative review of wind farm control with the objective of
  wind farm power maximization}.
\newblock {\em Journal of Wind Engineering {\&} Industrial Aerodynamics},
  192(May):45--73, 2019.

\bibitem{Kheirabadi2020}
Ali~C. Kheirabadi and Ryozo Nagamune.
\newblock {Real-time relocation of floating offshore wind turbine platforms for
  wind farm efficiency maximization: An assessment of feasibility and
  steady-state potential}.
\newblock {\em Ocean Engineering}, 208(May), 2020.

\bibitem{Mayne2000}
D.Q. Mayne, J.B. Rawlings, C.V. Rao, and P.O.M. Scokaert.
\newblock {Constrained model predictive control: Stability and optimality}.
\newblock {\em Automatica}, 36(6):789--814, 2000.

\bibitem{Mayne2014}
David~Q. Mayne.
\newblock {Model predictive control: Recent developments and future promise}.
\newblock {\em Automatica}, 50(12):2967--2986, 2014.

\bibitem{Ellis2014}
Matthew Ellis, Helen Durand, and Panagiotis~D. Christofides.
\newblock {A tutorial review of economic model predictive control methods}.
\newblock {\em Journal of Process Control}, 24(8):1156--1178, 2014.

\bibitem{Muller2017}
Matthias~A. Muller and Frank Allgower.
\newblock {Economic and Distributed Model Predictive Control : Recent
  Developments in Optimization-Based Control}.
\newblock {\em SICE Journal of Control, Measurement, and System Integration},
  10(2):39--52, 2017.

\bibitem{Liu2012a}
Jinfeng Liu, Xianzhong Chen, David~Mu{\~{n}}oz {De La Pe{\~{n}}a}, and
  Panagiotis~D. Christofides.
\newblock {Iterative distributed model predictive control of nonlinear systems:
  Handling asynchronous, delayed measurements}.
\newblock {\em IEEE Transactions on Automatic Control}, 57(2):528--534, 2012.

\bibitem{Kuwata2007}
Yoshiaki Kuwata, Arthur Richards, Tom Schouwenaars, and Jonathan~P. How.
\newblock {Distributed robust receding horizon control for multivehicle
  guidance}.
\newblock {\em IEEE Transactions on Control Systems Technology},
  15(4):627--641, 2007.

\bibitem{Richards2007}
A~Richards and J~P How.
\newblock {Robust distributed model predictive control}.
\newblock {\em International Journal of Control}, 80(9):1517--1531, 2007.

\bibitem{Maestre2011}
J.~M. Maestre, D.~{Mu{\~{n}}oz De La Pe{\~{n}}a}, E.~F. Camacho, and T.~Alamo.
\newblock {Distributed model predictive control based on agent negotiation}.
\newblock {\em Journal of Process Control}, 21(5):685--697, 2011.

\bibitem{Muller2012a}
Matthias~A. M{\"{u}}ller, Marcus Reble, and Frank Allg{\"{o}}wer.
\newblock {Cooperative control of dynamically decoupled systems via distributed
  model predictive control}.
\newblock {\em International Journal of Robust and Nonlinear Control},
  22:1376--1397, 2012.

\bibitem{Stewart2011}
Brett~T. Stewart, Stephen~J. Wright, and James~B. Rawlings.
\newblock {Cooperative distributed model predictive control for nonlinear
  systems}.
\newblock {\em Journal of Process Control}, 21(5):698--704, 2011.

\bibitem{Liu2019a}
Peng Liu, Arda Kurt, and Umit Ozguner.
\newblock {Distributed Model Predictive Control for Cooperative and Flexible
  Vehicle Platooning}.
\newblock {\em IEEE Transactions on Control Systems Technology},
  27(3):1115--1128, 2019.

\bibitem{Pannek2013}
J{\"{u}}rgen Pannek.
\newblock {Parallelizing a state exchange strategy for noncooperative
  distributed NMPC}.
\newblock {\em Systems and Control Letters}, 62(1):29--36, 2013.

\bibitem{Asadi2018}
Elham~(Fatemeh) Asadi and Arthur Richards.
\newblock {Scalable distributed model predictive control for constrained
  systems}.
\newblock {\em Automatica}, 93:407--414, 2018.

\bibitem{Kheirabadi2020a}
Ali~C. Kheirabadi and Ryozo Nagamune.
\newblock {Distributed Economic Model Predictive Control -- Addressing
  Non-convexity Using Social Hierarchies}.
\newblock 2020.

\bibitem{Wang2017a}
Ruigang Wang, Ian~R. Manchester, and Jie Bao.
\newblock {Distributed economic MPC with separable control contraction
  metrics}.
\newblock {\em IEEE Control Systems Letters}, 1(1):104--109, 2017.

\bibitem{Lee2011}
Jaehwa Lee and David Angeli.
\newblock {Cooperative distributed model predictive control for linear plants
  subject to convex economic objectives}.
\newblock In {\em Proceedings of the IEEE Conference on Decision and Control},
  number~6, pages 3434--3439. IEEE, 2011.

\bibitem{Lee2012}
Jaehwa Lee and David Angeli.
\newblock {Distributed cooperative nonlinear economic MPC}.
\newblock In {\em Proceedings of the 20th International Symposium on
  Mathematical Theory of Networks and Systems (MTNS)}, 2012.

\bibitem{Driessen2012a}
P.~A.A. Driessen, R.~M. Hermans, and P.~P.J. {Van Den Bosch}.
\newblock {Distributed economic model predictive control of networks in
  competitive environments}.
\newblock In {\em Proceedings of the IEEE Conference on Decision and Control},
  pages 266--271, 2012.

\bibitem{Chen2012}
Xianzhong Chen, Mohsen Heidarinejad, Jinfeng Liu, and Panagiotis~D.
  Christofides.
\newblock {Distributed economic MPC: Application to a nonlinear chemical
  process network}.
\newblock {\em Journal of Process Control}, 22(4):689--699, 2012.

\bibitem{Albalawi2017a}
Fahad Albalawi, Helen Durand, and Panagiotis~D. Christofides.
\newblock {Distributed economic model predictive control with Safeness-Index
  based constraints for nonlinear systems}.
\newblock {\em Systems and Control Letters}, 110:21--28, 2017.

\bibitem{Wolf2012}
Inga~J. Wolf, Holger Scheu, and Wolfgang Marquardt.
\newblock {A hierarchical distributed economic NMPC architecture based on
  neighboring-extremal updates}.
\newblock In {\em Proceedings of the American Control Conference}, pages
  4155--4160. IEEE, 2012.

\bibitem{Kohler2018b}
Philipp~N. K{\"{o}}hler, Matthias~A. M{\"{u}}ller, and Frank Allg{\"{o}}wer.
\newblock {A distributed economic MPC framework for cooperative control under
  conflicting objectives}.
\newblock {\em Automatica}, 96:368--379, 2018.

\bibitem{Kheirabadi2020b}
Ali~C. Kheirabadi and Ryozo Nagamune.
\newblock {A Dynamic Parametric Wind Farm Model for Simulating Time-varying
  Wind Conditions and Floating Platform Motion}.
\newblock 2020.

\bibitem{Robertson2014}
A.~Robertson, J.~Jonkman, and M.~Masciola.
\newblock {Definition of the Semisubmersible Floating System for Phase II of
  OC4}.
\newblock Technical report, National Renewable Energy Laboratory. Report
  number: NREL/TP-5000-60601, 2014.

\bibitem{Burton2011}
Tony Burton, Nick Jenkins, David Sharpe, and Ervin Bossanyi.
\newblock {\em {Wind Energy Handbook}}.
\newblock John Wiley {\&} Sons, Ltd, 2nd edition, 2011.

\bibitem{Bastankhah2016}
Majid Bastankhah and Fernando Port{\'{e}}-Agel.
\newblock {Experimental and theoretical study of wind turbine wakes in yawed
  conditions}.
\newblock {\em Journal of Fluid Mechanics}, 806:506--541, 2016.

\bibitem{Jimenez2009}
{\'{A}}ngel Jim{\'{e}}nez, Antonio Crespo, and Emilio Migoya.
\newblock {Application of a LES technique to characterize the wake deflection
  of a wind turbine in yaw}.
\newblock {\em Wind Energy}, 13(6):559--572, 2009.

\bibitem{Katic1986}
I.~Kati{\'{c}}, J.~H{\o}jstrup, and N.O. Jensen.
\newblock {A simple model for cluster efficiency}.
\newblock {\em Proceedings of the European Wind Energy Association Conference
  and Exhibition}, pages 407--410, 1986.

\bibitem{Jonkman2009}
J.~Jonkman, S.~Butterfield, W.~Musial, and G.~Scott.
\newblock {Definition of a 5-MW Reference Wind Turbine for Offshore System
  Development}.
\newblock Technical report, National Renewable Energy Laboratory. Report
  number: NREL/TP-500-38060, 2009.

\bibitem{Chisci2001}
L.~Chisci, J.~A. Rossiter, and G.~Zappa.
\newblock {Systems with persistent disturbances: Predictive control with
  restricted constraints}.
\newblock {\em Automatica}, 37(7):1019--1028, 2001.

\bibitem{Knudsen2015}
Torben Knudsen, Thomas Bak, and Mikael Svenstrup.
\newblock {Survey of wind farm control-power and fatigue optimization}.
\newblock {\em Wind Energy}, 18(8):1333--1351, 2015.

\end{thebibliography}

\end{document}